# Effects of Lévy Noise on a bistable Duffing System

Yong Xu[†1], Juanjuan Li[1], Jing Feng[1]

*Department of Applied Mathematics, Northwestern Polytechnical University, Xi'an 710072, China*

***Abstract:***

This paper numerically investigates the mean first passage time (MFPT) and phase transition of a bistable Duffing system driven by Lévy stable noise, which can reduce to the common Gaussian noise with the stability index 2. We obtain the stationary probability density functions by using Monte Carlo method to describe the change in distribution law and the influences of Lévy noise parameters on the stationary probability densities are discussed. The results indicate that the stability index, noise intensity and skewness parameter of Lévy noise can induce the phase transition behaviors (namely, the qualitative change of stationary probability density function). Furthermore, the MFPT is calculated as functions of different parameters, which implies that the effects of the stability index, noise intensity and the skewness parameter on MFPT are quite different, and apparent distinctions between Lévy and Gaussian noise can be observed.

**Key words:** Lévy noise, bistable Duffing system, stationary probability density, phase transition, mean first passage time (MFPT).

**PACS numbers**: 05.40.Fb, 02.70.Uu, 05. 40-a, 05.70.Fh

## I. Introduction

Stochastic dynamical systems arise as mathematical models for complex phenomena in almost areas (e.g., physics, chemistry, engineering and biology) under random fluctuations [1], and then the investigation of the influences of random forces (noises) is very important and becomes one of the most relevant and intensively developing research directions [2]. So far, there are many efficient tools to quantify stochastic dynamics, such as the stochastic resonance (SR) [3,4], the nonequilibrium noise-induced transition which may be phase transition or not [4, 5], the mean first passage time (MFPT) of a particle driven by noise over a fluctuating potential barrier (also called the noise-induced escape problem) [6,7], reentrance phenomenon,

[†] Corresponding author.
TEL/FAX: 86 29 8843 1657. E-mail: hsux3@nwpu.edu.cn

coherence resonance, thermal ratchet and so on[8].

In most of the existing theoretical studies, phase transition and MFPT often can be used to quantify the transient properties of the escape process in nonlinear dynamical systems, and noise-induced escape problem appears in many different fields and has been concentrated by a number of investigators [8-14]. Ghosh and Barik [8] examined the noise-induced transition in a fluctuating bistable potential of a driven quantum system and found the characteristic stationary probability distribution functions undergoing transition from unimodal to bimodal nature. Xu and He et al. [12] investigated the MFPT in a biased mono-stable system. Chiarella et al. [13] analyzed the D-bifurcation and P-bifurcation behavior of speculative financial markets. Xu and Gu [14] explored the stochastic bifurcations of a Duffing-Van der Pol oscillator with colored noise, and indicated the variation of system’s parameters, noise intensity and correlation time all can induce the stochastic bifurcations.

Most of authors fastened their attentions on Gaussian case only. However, Gaussian noise is just an ideal model and can not describe random fluctuations with large jumps. In fact, non-Gaussian noise has been widely observed in various areas such as physics, biology, seismology, electrical engineering and finance [1]. Lévy noise is a class of non-Gaussian noise, and exhibits long heavy tails of the distribution which makes the samples paths discontinuous in time. Compared with Gaussian noise, Lévy noise is more frequently encountered in nature, since long jumps are associated with a complex structure of the environment [15]. Nowadays, the applications of Lévy noise can be found in description of the dynamics in plasmas, diffusion in the energy space, self-diffusion in micelle system, and transport in polymer systems under conformational motion to the spectral analysis of paleoclimatic or economic data [16]. Although the transient properties of the escape process induced by Lévy noise have been largely explored in various first-order dynamical systems [1, 15, 17-23], little has been done for second-order Duffing system [8], and the case of bistable Duffing systems taking into account its full inertial dynamics has hitherto not yet been considered. So it is worth studying the phase transition and MFPT induced by Lévy noise in a Duffing systems.

This paper aims to analyze the effects of Lévy parameters on the phase transition and MFPT in a bistable Duffing system. First, the stationary probability distribution functions are calculated by using the Monte Carlo method, and two types qualitative changes in unimodal and bimodal are observed. The results indicate that the stability index, noise intensity and skewness parameter of Lévy noise can change the peak's number of the stationary probability distribution functions, and then the phase transition behavior is observed. Moreover the bifurcation diagrams of the system in corresponding parameter plane are presented. Finally the MFPT is discussed for different parameters, which shows the effects of the stability index and that of skewness parameter are completely different.

The structure of the paper is as follows: In section II, Duffing system driven by Lévy noise is introduced. The density functions of Lévy distribution are investigated in two different situations. In section III, the stationary probability densities and the corresponding bifurcation diagrams of system are obtained in different parameter planes, from which the phase transition behaviors are discussed. In section IV, the MFPT functions with different parameters (i.e. the stability index, noise intensity and skewness parameter) are examined, and the relevant interpretations are given. In section V, the paper is ended with conclusions and remarks.

## II. A Duffing system with Lévy noise

Consider a bistable Duffing system [3,24,25] driven by Lévy noise, which can be described in terms of the equation

$$\ddot{x} = -\gamma \dot{x} - \frac{dV(x)}{dx} + \eta(t), \tag{1}$$

where $\gamma$ is the damping parameter, $V(x)$ the potential function defined as

$$V(x) = -x^2/2 + x^4/4 \tag{2}$$

with two stable fixed points $x_{s1} = -1, x_{s2} = 1$ and one unstable point $x_{un} = 0$, $\eta(t)$ denotes the Lévy noise with stability index $\alpha$ $(0 < \alpha \le 2)$. Note that $\eta(t)$ is the formal time derivative of a Lévy process $\zeta(t)$, which can be viewed as a generalized wiener process, obeying Lévy distribution $L_{\alpha,\beta}(\zeta;\sigma,\mu)$ with characteristic function [26], $\Phi(k) = \int_{-\infty}^{+\infty} d\zeta e^{-ik\zeta} L_{\alpha,\beta}(\zeta;\sigma,\mu)$.

Furthermore,

$$\alpha \in (0,1) \cup (1,2], \qquad \Phi(k) = \exp\left[ i\mu k - \sigma |k| \left( 1 + i\beta sgn(k) \frac{2}{\pi} \ln|k| \right) \right], \tag{3}$$

$$\alpha = 1, \qquad \Phi(k) = \exp\left[ i\mu k - \sigma^{\alpha} |k|^{\alpha} \left( 1 - i\beta sgn(k) \tan\frac{\pi\alpha}{2} \right) \right], \quad (4)$$

Here $\alpha(\alpha \in (0,2])$ denotes the stability index, and the parameter $\beta$ $(\beta \in [-1,1])$ represents the skewness parameter, $\sigma$ $(\sigma \in (0,\infty))$ and $\mu$ $(\mu \in R)$ are the scale and mean parameter respectively, and $D = \sigma^{\alpha}$ is the noise intensity. From Eq.(3) we know the Lévy distribution will degenerate into a Gaussian distribution for $\alpha = 2$ (and an arbitrary skewness parameter $\beta$). In Fig.1, the Lévy density functions $L_{\alpha,\beta}(\zeta;\sigma,\mu)$ under different stability indexes and skewness parameters are simulated by using Janicki-Weron algorithm [16,26] with the parameters given as $\mu = 0.0, \sigma = 1.0$. It illustrates that the Lévy density function is symmetric for $\beta = 0$, and for $\alpha < 1, L_{\alpha,\beta}(\zeta;\sigma,\mu)$ shows left-skewed with $\beta < 0$ and right-skewed with $\beta > 0$, and for $\alpha > 1, L_{\alpha,\beta}(\zeta;\sigma,\mu)$ shows right-skewed with $\beta < 0$ and left-skewed with $\beta > 0$. Most importantly, the positions of the maxima of $L_{\alpha,\beta}(\zeta;\sigma,\mu)$ is different for $\alpha < 1$ and $\alpha > 1$ for the same skewness parameter, therefore the variation of the stability index $\alpha$ (from $\alpha < 1$ to $\alpha > 1$ ) may lead to the phase transition and different changes of the MFPT in system (1), and while $\alpha$=1, $L_{\alpha,\beta}(\zeta;\sigma,\mu)$ is the Cauchy case which will be considered in detail.

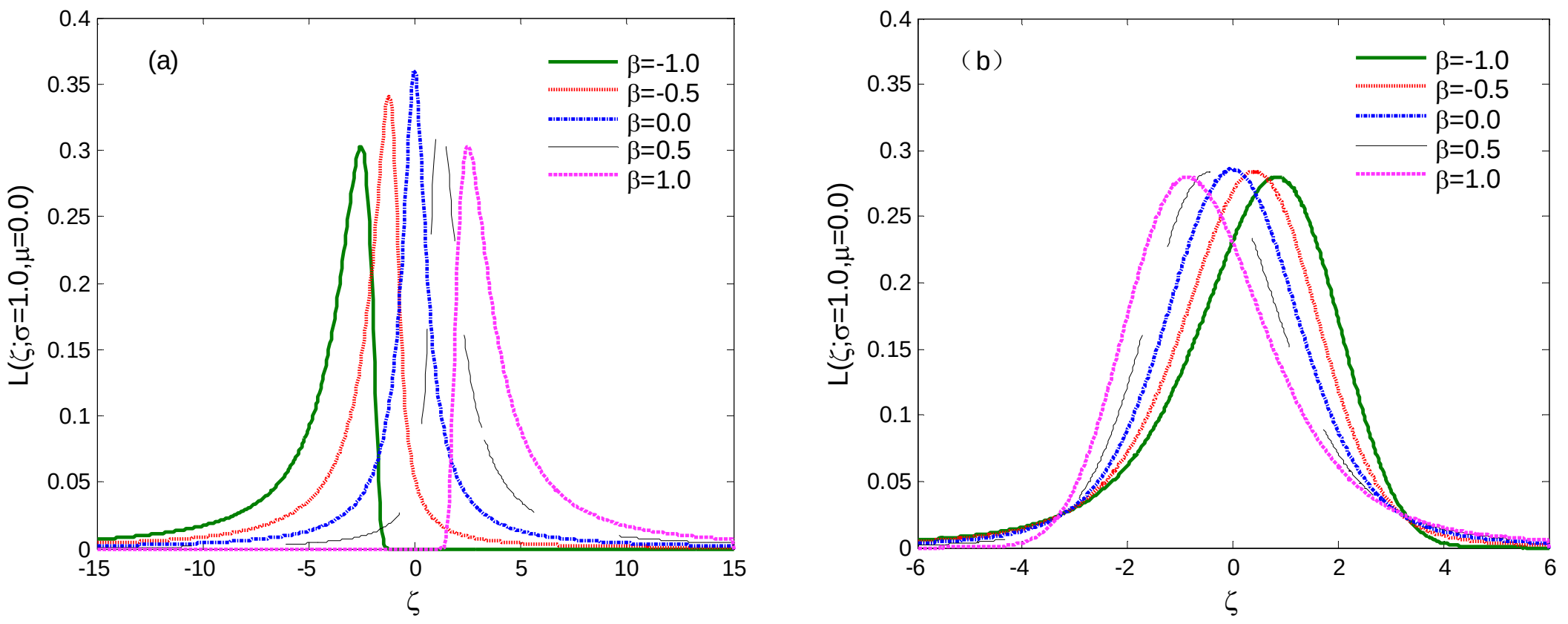

Fig.1. Lévy density functions $L_{\alpha,\beta}(\zeta;\sigma,\mu)$ with different skewness parameters for (a) $\alpha = 0.8$, (b) $\alpha = 1.6$.

# III. The phase transition induced by Lévy Noise

This section is to cope with the effects Lévy Noise on phase transition behaviors by analyzing the qualitative changes of the shape of the stationary probability density, namely, if the peak's number of the stationary probability distribution functions changes, such as two types qualitative change in unimodal and bimodal, then the phase transition occurs. Here the Monte Carlo method is employed to calculate the joint stationary probability density, and the corresponding stationary marginal probability density of random state variable $x(t)$ is acquired as well. With the help of the stationary marginal probability density, the behaviors of the phase transition are discussed with different noise parameters. In the following context we define the damping and mean parameter as $\gamma = 0.75$ and $\mu = 0$ respectively.

In order to obtain the probability density of stochastic state variable $x(t)$, letting $x = x$ and $\dot{x} = y$, after that the system (1) can be replaced by two first-order equations

$$\begin{cases} \dot{x} = y, \\ \dot{y} = -\gamma y - \dfrac{dV(x)}{dx} + \eta(t), \end{cases} \tag{5}$$

and the relative discrete form can be taken as [4,20]

$$\begin{cases} x_{n+1} = x_n + y_n \Delta t, \\ y_{n+1} = y_n + \left[ -\gamma y_n + x_n - x_n^{\,3} \right] \cdot \Delta t + \Delta t^{1/\alpha} \zeta, \end{cases} \tag{6}$$

where $\zeta$ denotes Lévy distributed random number with stability index $\alpha$ and noise intensity $D$. Then the fourth-order Runge-Kutta method is used to get the numerical solution of Eq.(6) and the joint stationary probability density functions of the state variables $x(t)$ and $\dot{x}(t)$ are obtained. All our simulations are performed with a time step $\Delta t = 0.01$ and the initial values $x(0) = \dot{x}(0) = 0.01$.

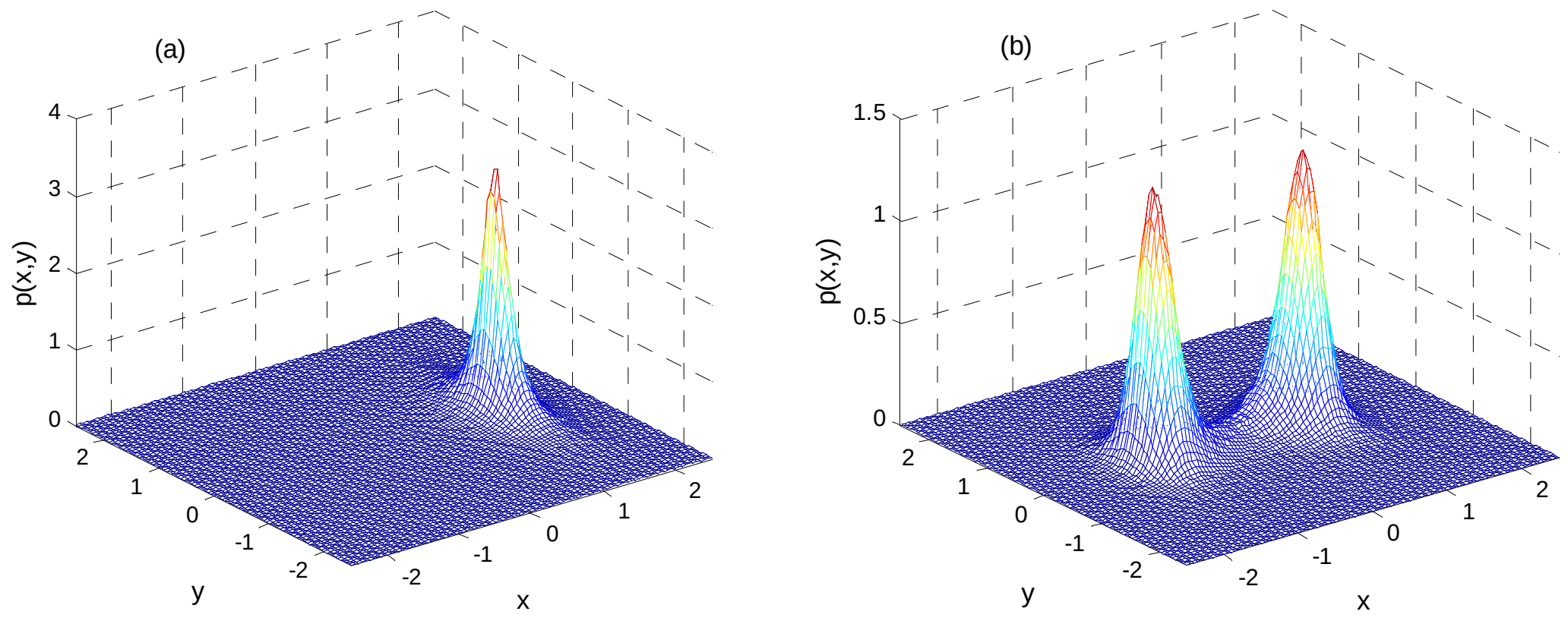

Fig.2. The joint stationary probability density functions of $x$ and $\dot{x}$ with different parameters of Lévy noise (a) $\alpha = 0.93, \beta = 0.9, D = 0.2$. (b) $\alpha = 1.5, \beta = 0.1, D = 0.1$.

Depending on above approach, the joint stationary probability density functions

$p(x,y)$ are simulated, as shown in Fig.2. The corresponding stationary marginal probability densities $p(x)$ are given in Fig.3. From Fig.2(a) to Fig.2(b), we can observe the shape's changes of $p(x,y)$, namely, as $\alpha = 0.93, \beta = 0.9, D = 0.2$ only one peak in the curve of $p(x,y)$, and then when $\alpha = 1.5, \beta = 0.1, D = 0.1$, two peaks appears in Fig.2(b), that implies the adjustment of Lévy noise parameters can cause the phase transition behaviors in the system (1) , and this phase transition behaviors (from unimodal to bimodal) can be observed more obviously in Fig.3.

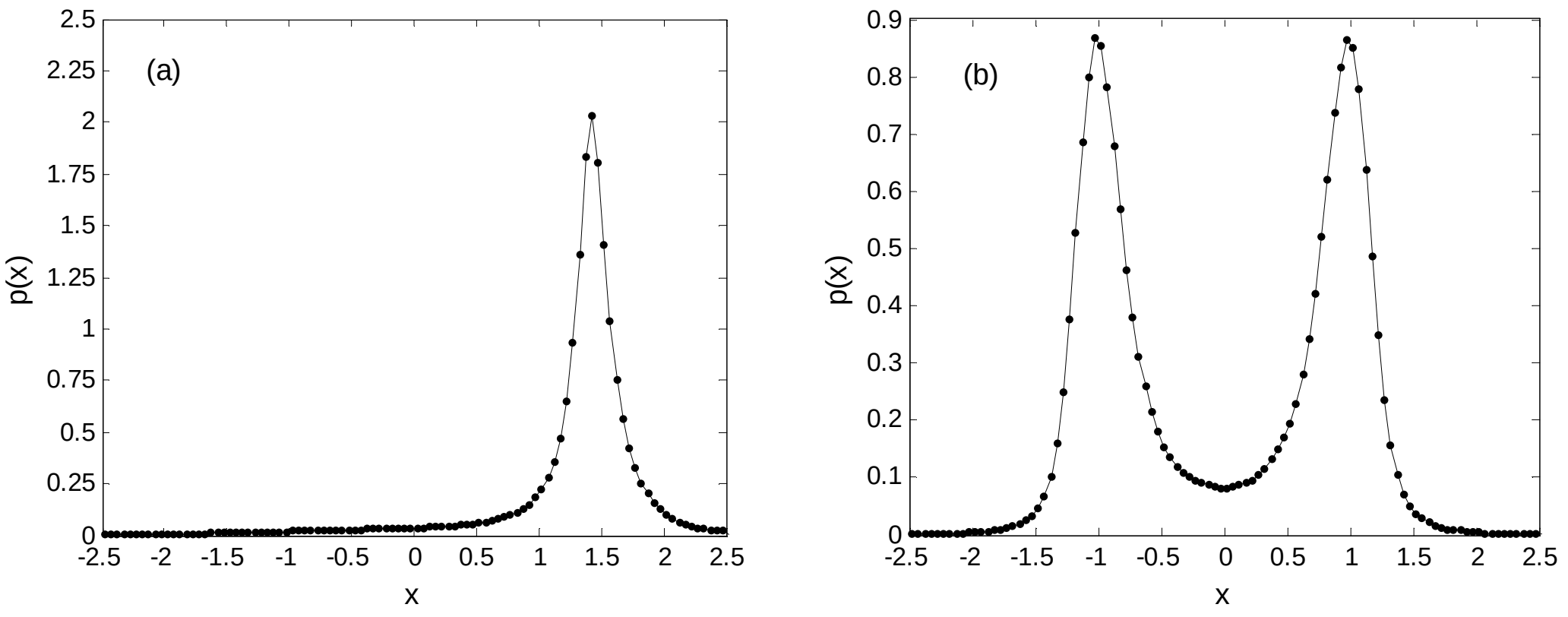

Fig.3. The corresponding stationary marginal probability density functions $p(x)$ with different parameters of Lévy noise (a) $\alpha = 0.93, \beta = 0.9, D = 0.2$ (b) $\alpha = 1.5, \beta = 0.1, D = 0.1$.

Next, we are devoted to discussing the phase transition behaviors of system (1) through qualitative changes of the stationary marginal probability density $p(x)$, where the number and the extreme of the stationary densities will be carefully examined. Note that the symmetry of Lévy density function for $\beta \in [0,1]$ and $\beta \in [-1,0]$, we only consider the overall range of parameters $\beta \in [0,1]$ and $\alpha \in (0,2]$.

From the discussions of Lévy noise, we know that when $\beta > 0$ the Lévy density is right-skewed for $\alpha < 1$ and left-skewed for $\alpha > 1$ , thus we judge that the phase transition of system (1) will take place as the variation of stability index from $0 < \alpha < 1$ to $1 < \alpha \le 2$ . And the numerical evolution of corresponding trajectories is known to be instable for $\alpha = 1, \beta \ne 0$. So the case of $0 < \alpha < 1$ and $1 < \alpha \le 2$ with $\beta \ne 0$ are considered here, and the special situation of $\alpha = 1, \beta \ne 0$ is individually analyzed later. In Fig.4 the phase transition behaviors of system (1) with different parameters of Lévy noise are demonstrated. We choose the noise intensity $D = 0.1$ and examine the effects of stability index and skewness parameter on the stationary marginal probability density $p(x)$.

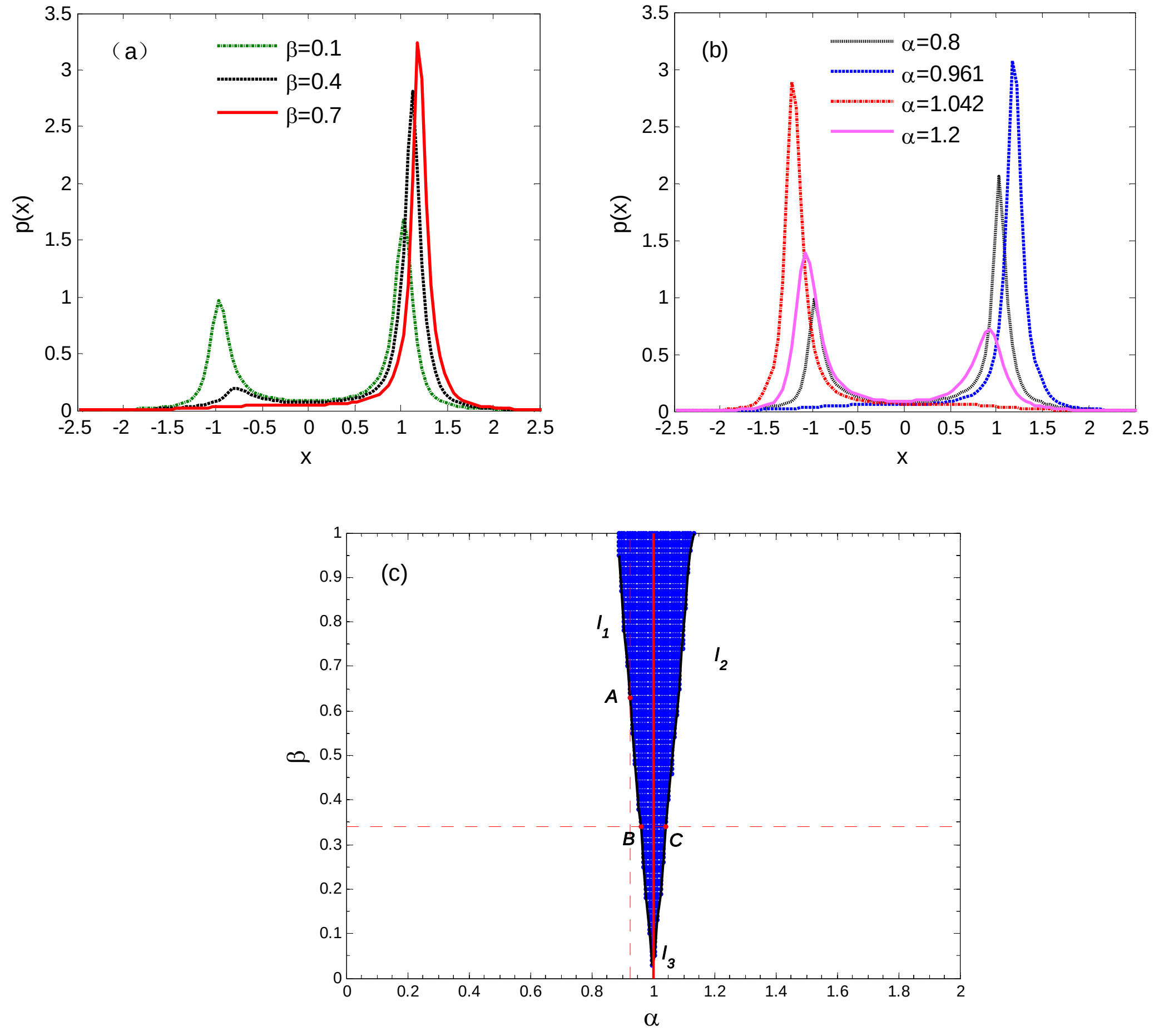

Fig.4. The stationary marginal probability densities $p(x)$ for (a) $\alpha = 0.925, D = 0.1$, (b) $\beta = 0.34, D = 0.1$ with different values of the skewness parameters and stability indexes, and the bifurcation diagram (c) of the system (1) in the parameter plane $(\alpha, \beta)$ for $D = 0.1$.

Fig.4.(a) shows the stationary marginal probability densities $p(x)$ for different values of the skewness parameters $\beta$ with $D = 0.1$, $\alpha = 0.925 < 1$, it can be obviously observed that the qualitative changes between unimodal and bimodal. For smaller $\beta$, the probability densities $p(x)$ has two peaks, and the positions of the peaks are near by two stable points of the potential function, however when increasing $\beta$ successively, the height of left peak gradually decreases and the right peak increases, then $p(x)$ changes from bimodal to unimodal, that illustrates the skewness parameter induces the phase transition of the system (1). According to Fig.4.(b), we discover that the parameter $\alpha$ plays an important role in phase transition. When noise intensity $D$ is fixed, $p(x)$ undergoes a succession of phase transition (bimodal-unimodal-bimodal) as $\alpha$ from $(0,1)$ to $(1,2]$, and the position of the unimodal stationary marginal probability density $p(x)$ is different for $\alpha < 1$

and $\alpha > 1$, that is because the differences in the positions of the maxima of the Lévy distribution for $\alpha < 1$ and $\alpha > 1$. In Fig.4.(c), the bifurcation diagram in the parameter plane $(\alpha, \beta)$ is presented, here the tinted region represents the unimodal distribution and colorless region is the bimodal distribution, and lines $l_1$, $l_2$ respectively correspond to the disappearance and appearance of one maxima of $p(x)$, point A is the intersection point of the vertical dash line $\alpha = 0.925$ and $l_1$, point B and C are the intersection point of the horizontal dash line $\beta = 0.34$ and $l_1$, $l_2$ respectively, line $l_3$ stands for the Cauchy case of $\alpha = 1.0$, and the detail discussions of which will be considered in Fig.6. It's apparent that the unimodal region gradually reduces and becomes narrower with the decrease of skewness parameter $\beta$, and when $\beta$ decreases to zero (i.e. the symmetric Lévy noise) with any $\alpha$, the stationary marginal probability densities $p(x)$ always have the symmetric and bimodal structure along with $x = 0$, this means the symmetric Lévy noise can't induce the phase transition with $D = 0.1$. More importantly, the skewness parameters $\beta$ also can't induce the phase transition in the system (1) driven by Gaussian noise ($\alpha$=2.0).

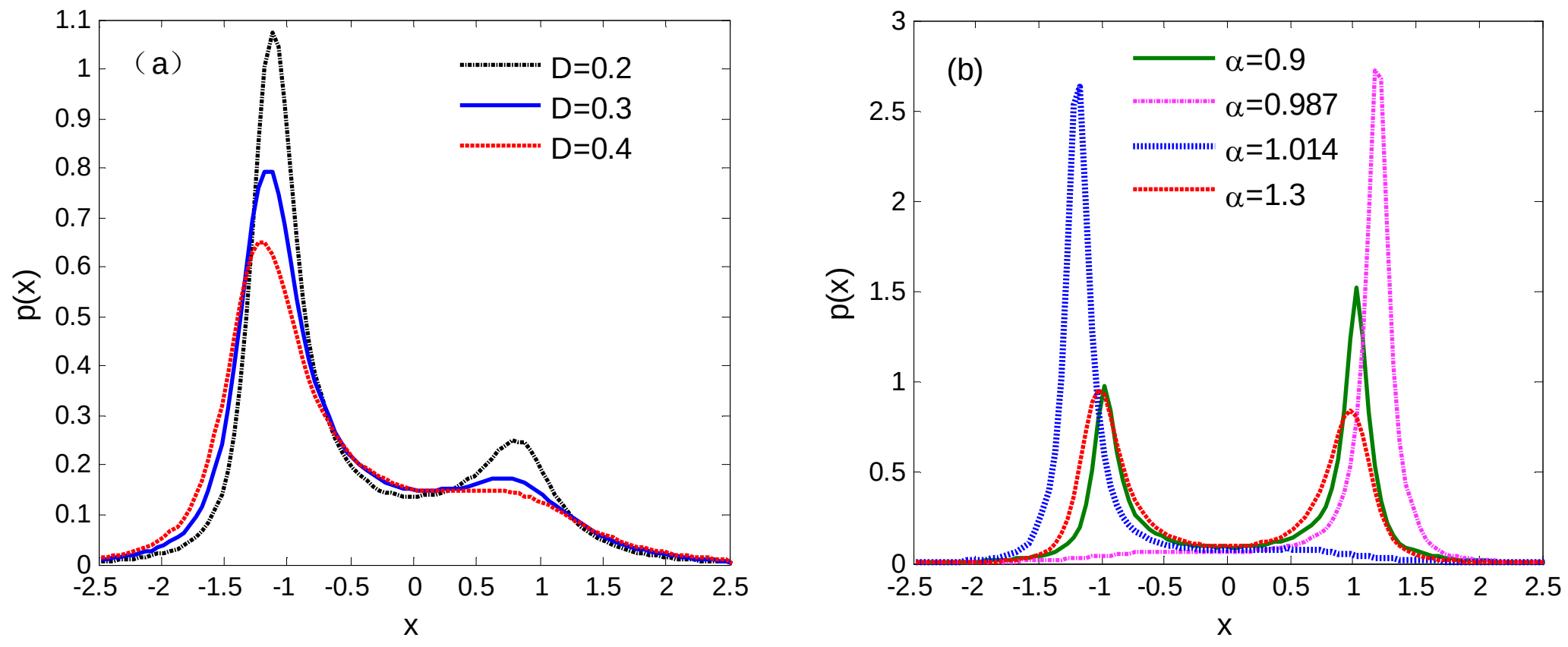

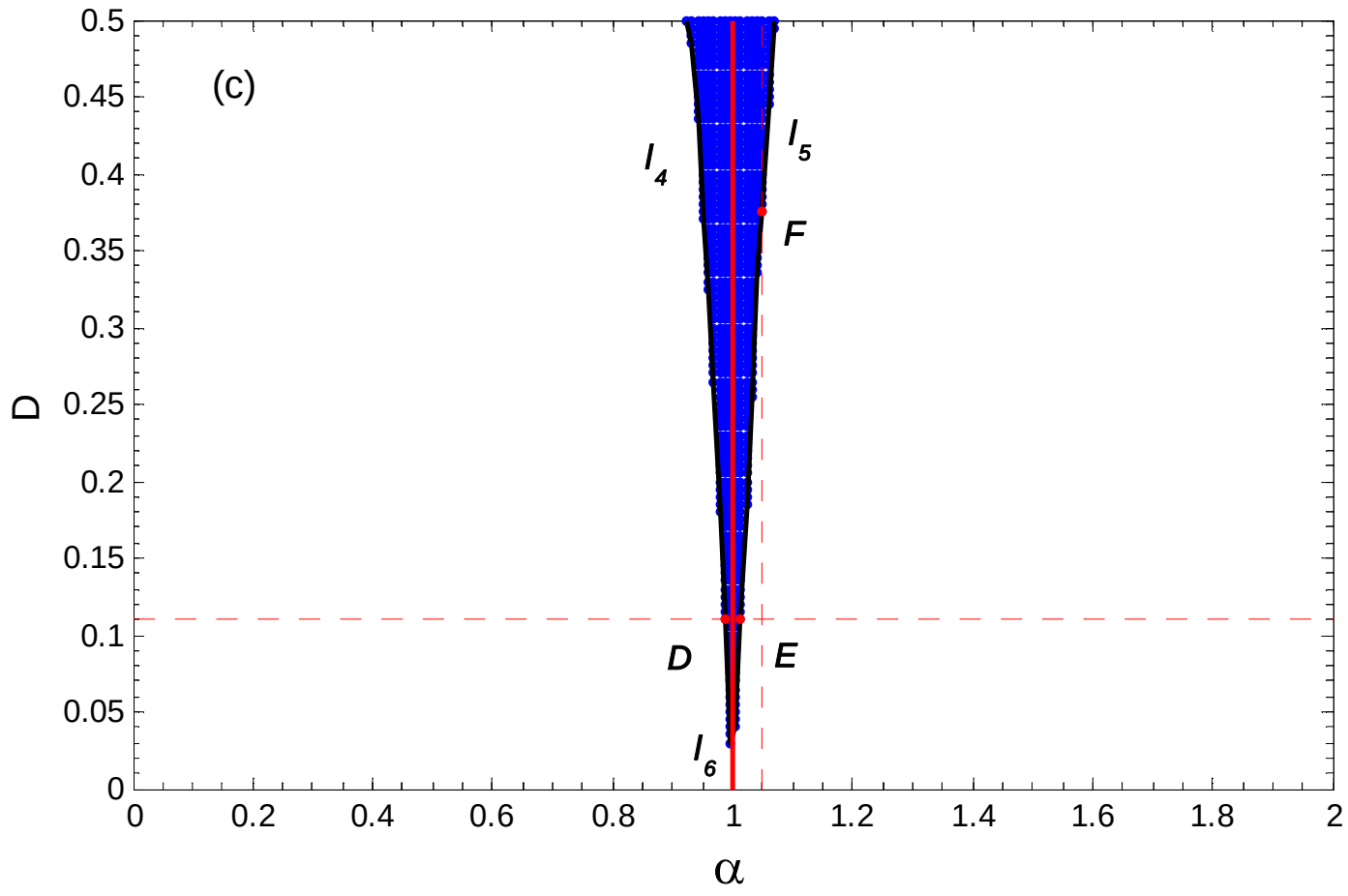

Fig.5. The stationary marginal probability densities $p(x)$ of the system (1) for (a) $\alpha=1.05, \beta=0.1$, (b) $\beta=0.1, D=0.11$ with the varying noise intensity $D$ and stability index $\alpha$, and the bifurcation diagram (c) of the system (1) in the parameter plane $(\alpha, D)$ with $\beta=0.1$.

From the discussions of Fig.4, we find the skewness parameter and stability index all can induce the phase transition of the system (1) for fixed noise intensity, and the stability index has a great effect on the phase transition. Similarly, the influence of the noise intensity and stability index on the phase transition with a fixed skewness parameter is examined in Fig.5. Fig.5. (a) shows the curves of the stationary marginal probability density $p(x)$ changes from two peaks to one peak with $D$ increases from 0.2 to 0.4, thus the noise intensity induces one-time phase transition, and Fig.5.(b) implies the enhancement of the stability index $\alpha$ induced two-time phase transition. In Fig.5.(c), the bifurcation diagram in the parameter plane $(\alpha, D)$ is given, where the tinted region describes the unimodal distributions and the gap region displays the bimodal distributions, lines $l_4$ and $l_5$ represent the critical values of $\alpha$ and $D$ of the phase transition, point F is the intersection point of the vertical dash line $\alpha=1.05$ and $l_5$, point D and E are the intersection point of the horizontal dash line $D=0.11$ and $l_4$, $l_5$ respectively, line $l_6$ is the Cauchy case ($\alpha$=1.0). As can be seen, the phase transition behaviors have been observed with $D$ or $\alpha$ changing between the gap and tinted region, and the bimodal region is increasing with decreasing of $D$. For $\alpha=2.0$, the stationary marginal probability densities $p(x)$ are bimodal structure throughout the changing of $D$, which illustrates the phase transition of the system (1) driven by Gaussian noise ($\alpha$=2.0) can't be induced by varying of $D$.

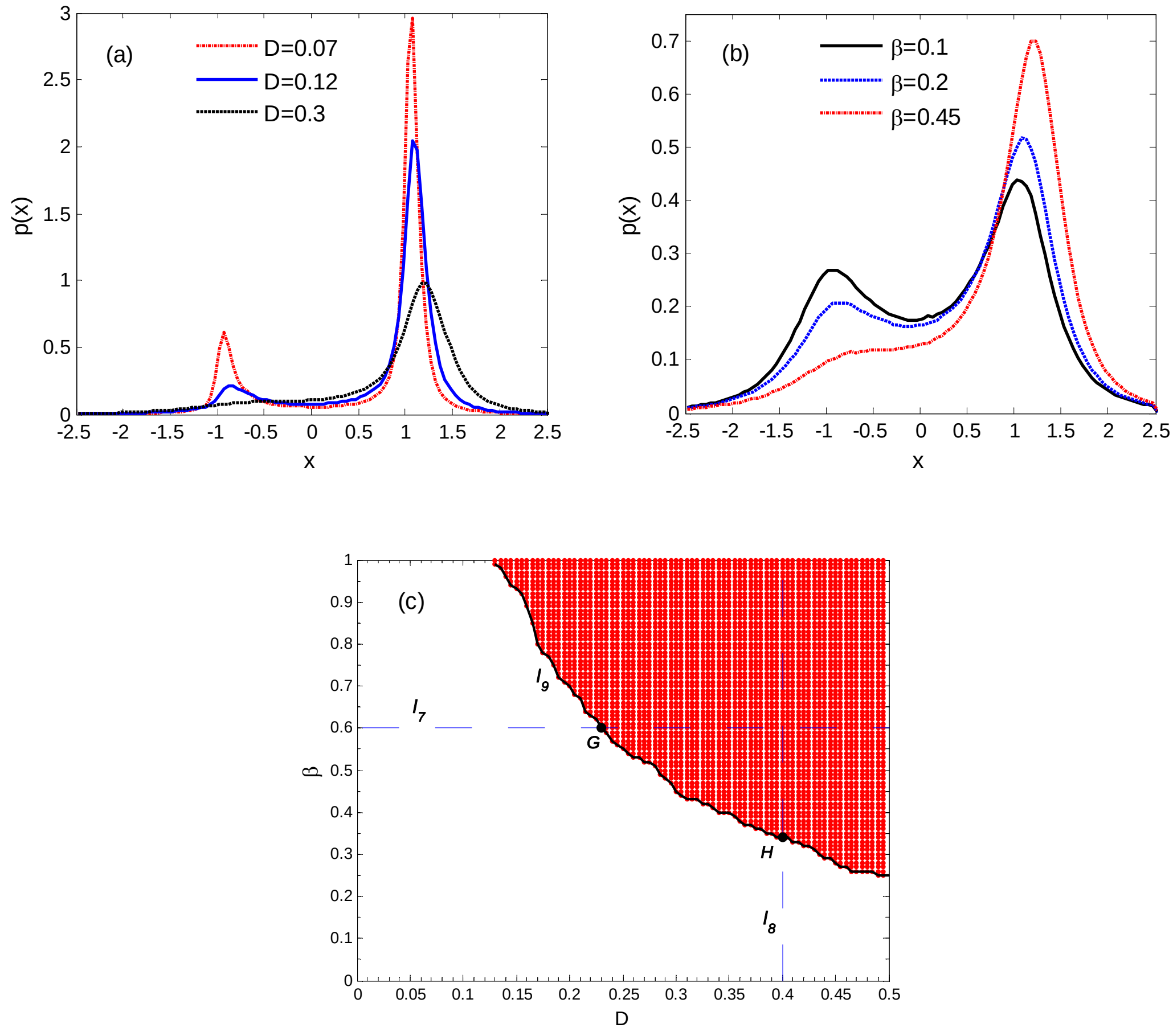

Fig.6. The stationary marginal probability densities $p(x)$ of the system (1) for $(a)$ $\alpha = 1.0, \beta = 0.6$ with varying noise intensity $D$, $(b)$ $\alpha = 1.0, D = 0.4$ with changing skewness parameter $\beta$, and the bifurcation diagram (c) in the parameter plane $(D, \ \beta)$ with $\alpha = 1.0$.

In Fig.6, the phase transition of system (1) is discussed with the special case of $\alpha = 1.0$, and the influences of noise intensity and skewness parameter on the stationary marginal probability densities $p(x)$ are investigated. Fig.6.(c) exhibits the corresponding bifurcation diagram in the parameter plane $(D, \beta)$ for $\alpha = 1.0$, line $l_9$ is the critical points of unimodal and bimodal structure, and the tinted region is unimodal distribution. Point H is the intersection point of the vertical dash line $D = 0.4$ and $l_9$, and point G is the intersection point of the horizontal dash line $\beta = 0.6$ and $l_9$. It's clearly observed that the increase of $D$ and $\beta$ leads to the bimodal region gradually decreases and unimodal region constantly extends. When $\beta$ decreases to $0.25$ even smaller, the stationary marginal probability densities all have two peaks with the changing of $D$ in $(0, 0.5)$. Fixed $\beta = 0.6$, the phase transition is discovered in Fig.6.(a) as increasing $D$, which means one-time phase transition happened in the system (1) and this result is coincide with Fig.6(c). Fig.6.(b) shows the stationary marginal probability densities $p(x)$ as varying skewness parameter $\beta$ for $D = 0.4$, and one-time phase transition from a unimodal to a bimodal distribution occurs. All these subfigures illustrate that the noise intensity and skewness parameter can induce the phase transition for $\alpha = 1.0$.

## IV. Mean First Passage Time

The analysis of the phase transition gives us a perspective from a dynamical system viewpoint by focusing on the stationary marginal probability density of the stochastic state variable $x(t)$. Besides, the MFPT is also an appropriate choice to describe the escape properties of the system with random perturbations. In this section we consider the problem of escaping through digital simulation, here the MFPT is defined by the first time of the particle jumps from one potential well to another (i.e. from $x_{s1}$ to $x_{s2}$ or from $x_{s2}$ to $x_{s1}$), we mainly explore the MFPT in two directions ($T_+(x_{s1} \to x_{s2})$ and $T_-(x_{s2} \to x_{s1})$), then the effects of noise intensity $D$, stability index $\alpha$ and skewness parameter $\beta$ on the MFPT are examined. The final results are presented by averaging over $2\times10^4$ realizations. Similarly, the special case of $\alpha$=1.0, $\beta \neq 0$ is considered independently.

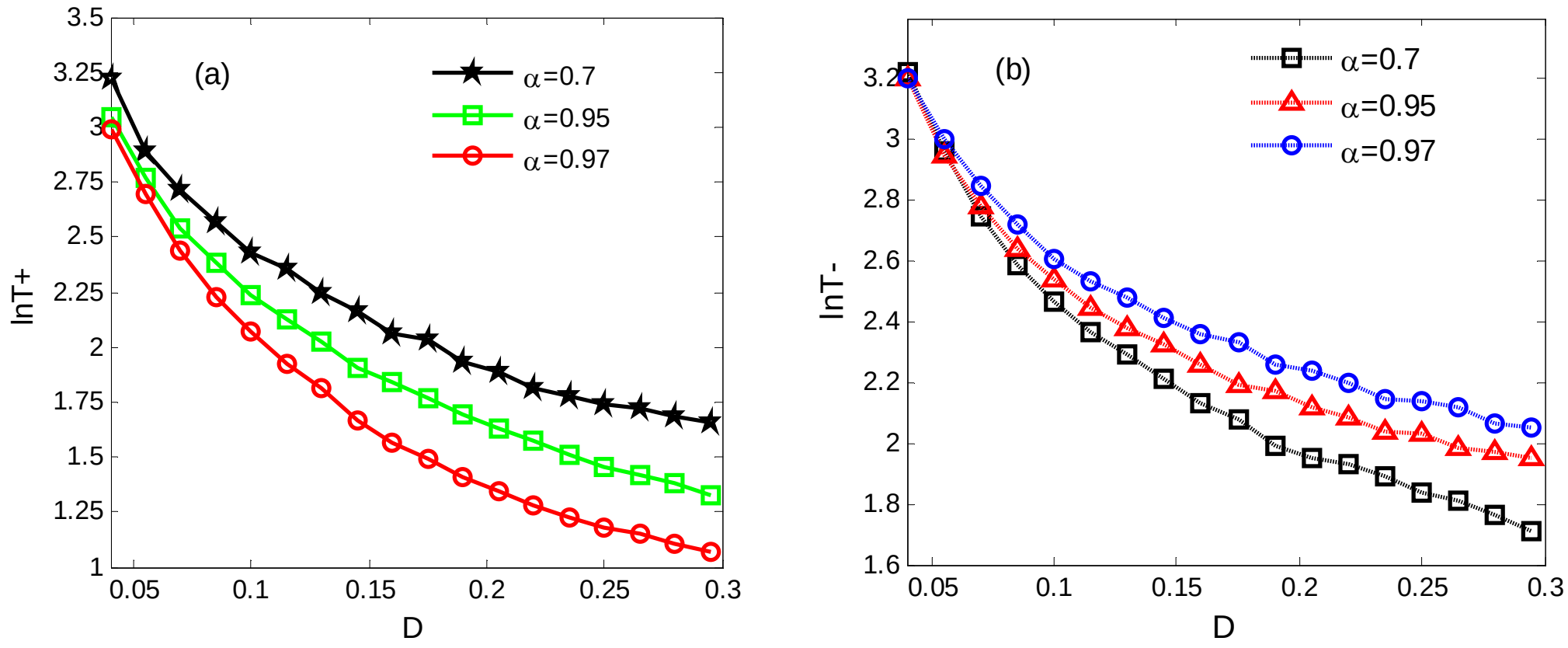

`````

```

Fig.7.The MFPT as functions of the noise intensity with different stability indexes $(a)$ $T_+(x_{s1} \to x_{s2})$ and $(b)$ $T_-(x_{s2} \to x_{s1})$. The value of the skewness parameter is $\beta = 0.1$.

According to the definition of MFPT, the MFPT functions of the system (1) from two opposite directions are presented in Fig.7 to Fig.10. In Fig.7, the MFPT functions versus the noise intensity with different stability indexes $\alpha$ are investigated. From which we observe that $T_+(x_{s1} \to x_{s2})$ and $T_-(x_{s2} \to x_{s1})$ present the decreasing and monotonic behavior with increasing of the noise intensity $D$, which means the increase of $D$ can speed up the transition of particle between $x_{s1}$ and $x_{s2}$, furthermore the changing range of $T_+(x_{s1} \to x_{s2})$ and $T_-(x_{s2} \to x_{s1})$ is different. For fixed $D$, the increase of $\alpha$ cause the decrease of $T_+(x_{s1} \to x_{s2})$ and increase of $T_-(x_{s2} \to x_{s1})$, thus illustrates that the effect of $\alpha$ on $T_+(x_{s1} \to x_{s2})$ and $T_-(x_{s2} \to x_{s1})$ is unlike for $\beta = 0.1$.

As can be intuitively expected, the skewness parameter $\beta$ similarly influences the MFPT of the system (1). Fig.8 shows the effects of $\beta$ on the $T_+(x_{s1} \to x_{s2})$ and $T_-(x_{s2} \to x_{s1})$ with different stability indexes $\alpha$. As can be seen in Fig.8, the function of $T_+(x_{s1} \to x_{s2})$ increases monotonously and $T_-(x_{s2} \to x_{s1})$ decreases monotonously as $\beta$ increases for $D = 0.1$, and the more stability indexes $\alpha$ approaches to 1.0, the faster MFPT increases or decreases. When noise is Gaussian case ($\alpha = 2.0$), our results suggest that MFPT is essentially independent of $\beta$.

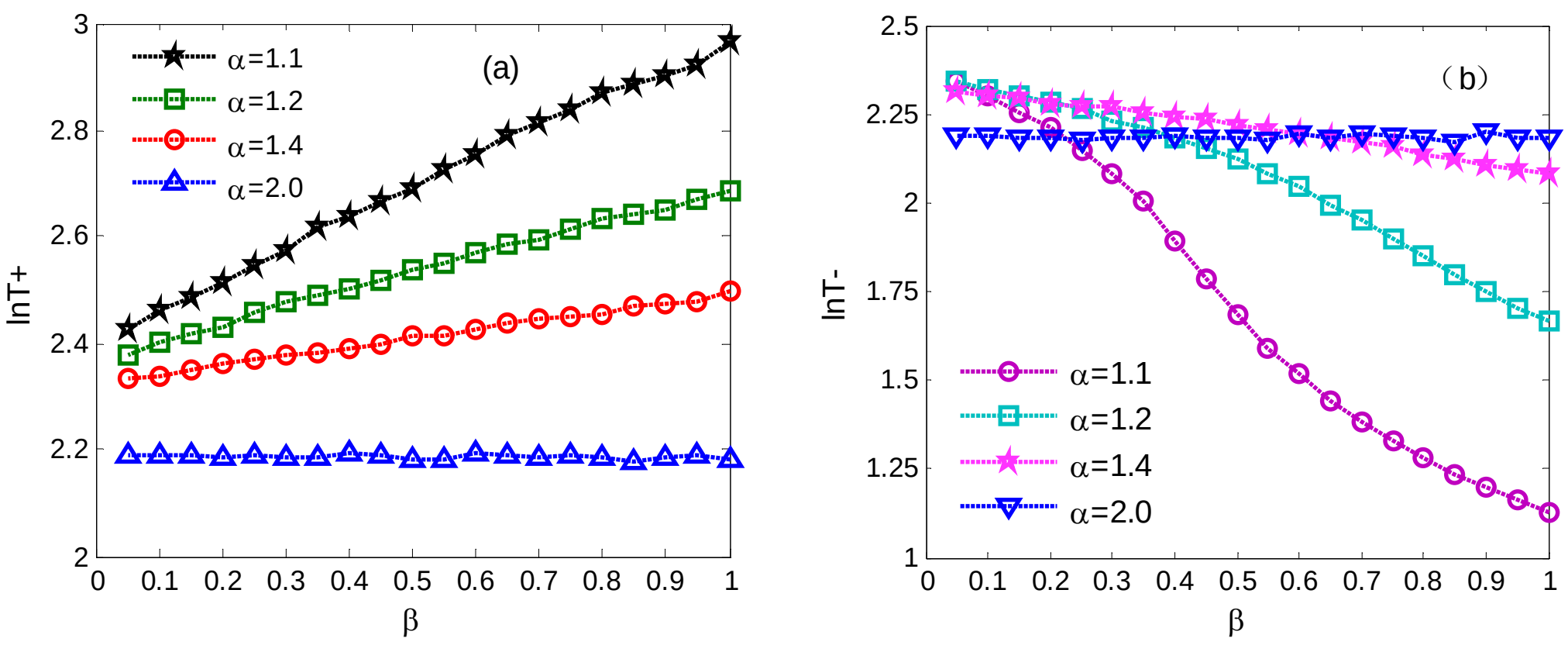

Fig.8. The MFPT versus the skewness parameters $\beta$ of Lévy noise for different stability indexes $\alpha$, (*a*) $T_+(x_{s1} \to x_{s2})$ and (*b*) $T_-(x_{s2} \to x_{s1})$. The noise intensity is $D = 0.1$.

When $\alpha = 1.0$, we independently consider the effects of noise intensity and skewness parameter on the MFPT in Fig.9, that illustrates the MFPT decreases monotonously with the increase of the noise intensity, and the values of $T_+(x_{s1} \to x_{s2})$ are much smaller than $T_-(x_{s2} \to x_{s1})$ with varying of $D$. Fig.9.(b) demonstrates the different influences of $\beta$ on $T_+(x_{s1} \to x_{s2})$ and $T_-(x_{s2} \to x_{s1})$, and one can find that the increase of $\beta$ can promote the transition of particle from $x_{s1}$ to $x_{s2}$, and restrain the jump of particle from $x_{s2}$ to $x_{s1}$.

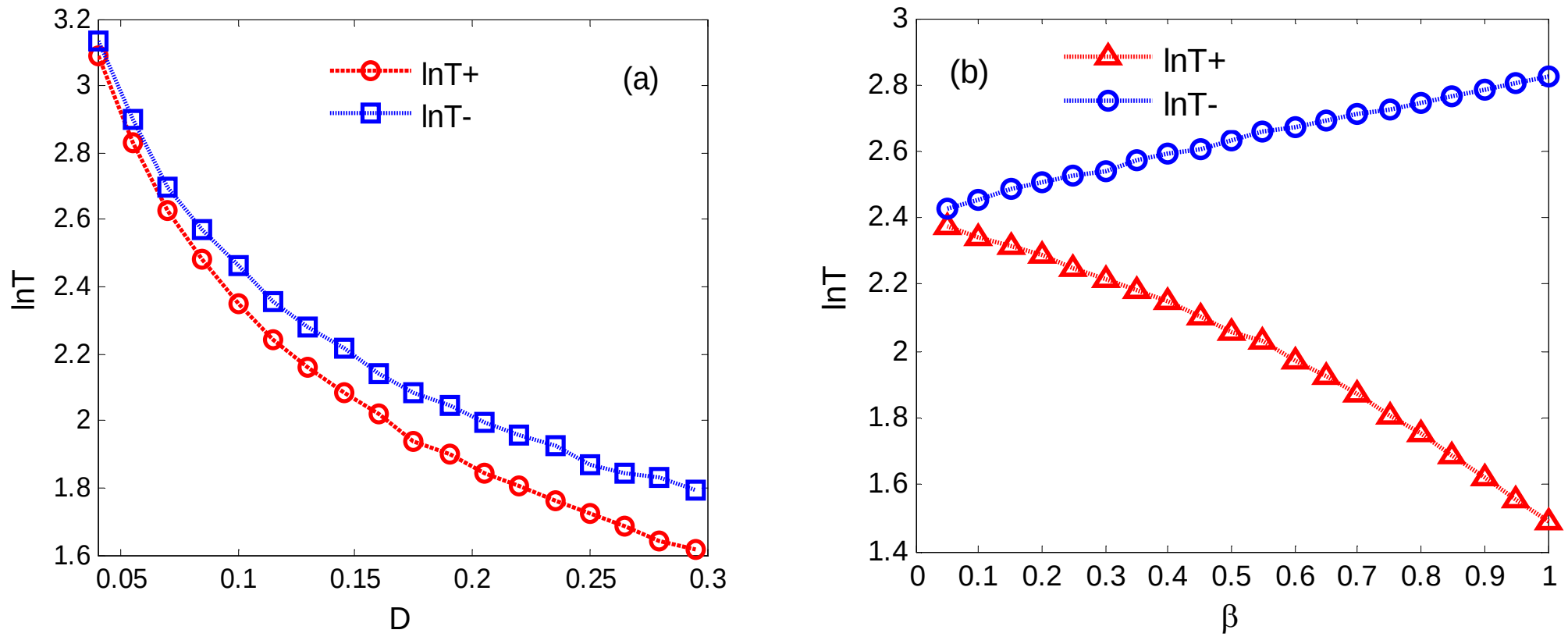

Fig.9. The MFPT as the functions of the noise intensities and skewness parameter for $(a)$ $\beta = 0.1$, $(b)$ $D = 0.1$ under the special case of $\alpha = 1.0$.

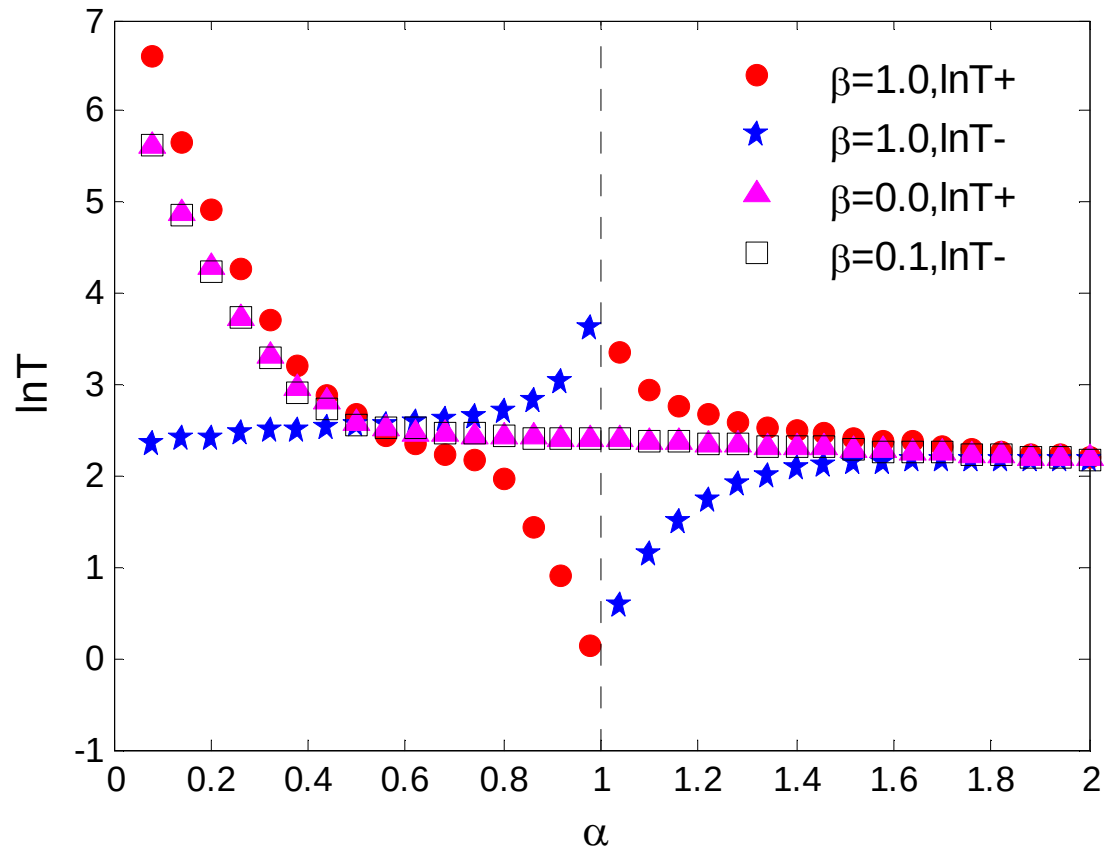

Fig.10. The MFPT as the function of the stability index of Lévy noise with fully asymmetric ($\beta$=1.0) and symmetric ($\beta$=0.0) case respectively. The value of the noise intensity is $D = 0.1$.

From above analysis of the MFPT in Fig.7, we find the stability index has a great effect on the MFPT, so that the MFPT as the functions of $\alpha$ with different skewness parameters are explored in Fig.10. For symmetric Lévy noise, the MFPT functions continuously decreases and this variation trend gradually becomes gentler with the increase of $\alpha$, meanwhile $T_+(x_{s1} \to x_{s2})$ and $T_-(x_{s2} \to x_{s1})$ almost have no difference for arbitrary $\alpha$. However, for asymmetric Lévy noise, the MFPT functions exhibit a discontinuous behavior at $\alpha = 1.0$, which is resulted from the discontinuous of the characteristic function of Lévy noise for $\alpha = 1.0, \beta \neq 0$. From Fig.10 we find the effects of $\alpha$ on $T_+(x_{s1} \to x_{s2})$ and $T_-(x_{s2} \to x_{s1})$ are very different with $\beta \neq 0$, and this result is consistent with the Fig.7.

## V. Conclusions

Founded on the numerical analysis of a bistable Duffing system driven by a non-Gaussian Lévy noise, the phase transition and MFPT are used to quantify the transient properties of the escape process. In this paper the effects of the parameters of Lévy noise on the phase transition and MFPT were carefully examined. And for the discontinuous Lévy characteristic function at $\alpha = 1.0, \beta \neq 0$, we respectively considered the case of stability index $\alpha = 1.0$ and $\alpha \neq 1.0$.

Originally, the phase transition behaviors of the system is discussed by two types of qualitative change of the stationary marginal probability densities obtained by Monte Carlo method, and the bifurcation diagrams in various parameter planes have been presented to show that the stability index, noise intensity and skewness parameter can induce the phase transition in a bistable Duffing system, but when stability index $\alpha = 2.0$ (i.e. the Lévy noise degenerates into the Gaussian case), the noise intensity can't induce the phase transition which is caused by the different mechanisms of Lévy noise and Gaussian noise .

Additionally, the MFPT as functions of the noise intensity, stability index and skewness parameter have been explored, and the results indicate that the increase of noise intensity can decrease the MFPT, and speed up the jump of particle between two potential wells. However, as the increase of stability index and skewness parameter, our findings implies that the effects of skewness parameter and stability index on MFPT of two directions are completely opposite. And when noise is Gaussian case ($\alpha = 2.0$), our results suggest that MFPT is essentially independent of skewness parameter. Meanwhile we discuss the MFPT in the case of the discontinuous Lévy characteristic function at $\alpha = 1.0, \beta \neq 0$.

In summary, our present study not only indicates the effects of noise intensity, stability index and skewness parameter of Lévy noise on the phase transition and MFPT but also demonstrates the significantly different effects of Lévy noise and Gaussian noise. A number of recent studies have shown that the Lévy noise is observed and concerned in many situations (e.g. biology and physics), and we hope that our present observations will be useful support to experimental findings.

## Acknowledgments

This work was supported by the NSF of China (Grant Nos. 10972181, 11102157),

NCET, the Shaanxi Project for Young New Star in Science & Technology and NPU Foundation for Fundamental Research, and SRF for ROCS, SEM. The authors thank the referees for their very valuable suggestions.